\documentclass{article}

\usepackage{spconf,amsmath,graphicx,hyperref}
\usepackage{tabularx}
\usepackage{booktabs}
\usepackage{multirow}
\usepackage{array}
\usepackage{pifont}
\usepackage[most]{tcolorbox}
\usepackage{xcolor}
\usepackage{orcidlink}
\usepackage{etoolbox}

\title{Beyond Latent Personalization: Integrating HRV-conditioned Personalization via LLM for Interpretable Stress Detection}
\name{
Wei-Hsiang Ma$^{1}$\,\orcidlink{0009-0002-6256-6336},
Ya-Tse Wu$^{2}$\,\orcidlink{0009-0009-0769-458X},
Hsing-Hang Chou$^{2}$\,\orcidlink{0009-0005-9806-2174},
Chi-Chun Lee$^{2}$\,\orcidlink{0000-0003-0186-4321}
}

\address{
$^{1}$Institute of Communication Engineering, National Tsing Hua University, Taiwan\\
$^{2}$Department of Electrical Engineering, National Tsing Hua University, Taiwan
}
\begin{document}
%
\maketitle
\begin{abstract}
Stress detection based on heart rate variability (HRV) has received increasing attention. However, existing personalized approaches integrate personal attributes with HRV through latent representations, limiting the interpretability of their interactions. In this work, we propose an LLM-enhanced personalized stress detection framework that integrates HRV features and personal attributes through LLM-based semantic interpretation. The proposed framework achieves a 29.9\% relative MCC improvement over the strongest baseline on TILES-2018 while providing human-readable descriptions of individual physiological responses. The results highlight the potential of LLM-based semantic interpretation for achieving both effective and interpretable personalized stress detection.
\end{abstract}
\vspace{-5mm}
\section{Introduction}
\vspace{-2mm}
\label{sec:intro}
With increasing awareness of mental health, physiological signal-based stress detection has become an important research topic. Among various physiological indicators, heart rate variability (HRV) has received considerable attention, as numerous studies have demonstrated its association with psychological stress \cite{immanuel2023heart, thielmann2022objective}. 
However, HRV responses vary across individuals due to personal attributes, such as age and gender \cite{valentini2009variables, koenig2016sex}, as well as work-related factors, such as shift schedules \cite{zhan2025factors}, suggesting that personal context should be considered when interpreting HRV patterns for stress detection.

Recent studies have incorporated personal attributes into HRV-based stress detection by aligning HRV with survey information through contrastive learning \cite{yang2025contrastive}, regularizing HRV embeddings using pairwise attribute similarities \cite{wang2024wild}, and modeling higher-order HRV--attribute interactions through hypergraph learning \cite{wang2025dynamic}. Despite their effectiveness, these feature-based approaches implicitly encode HRV--attribute interactions in latent representations, providing limited insight into how relatively static personal attributes contextualize dynamically changing HRV responses for stress detection \cite{bolpagni2024personalized}. Large language models (LLMs) offer a promising alternative, having been applied across medical domains to integrate clinically informative text into downstream tasks \cite{chen2025leveraging}. Recent efforts have translated and aligned non-linguistic, high-dimensional physiological signals with LLMs for various downstream tasks \cite{liu2024large,kim2024health,yang2025ecg}. Their ability to express contextual information provides a natural means of describing how personal attributes relate to physiological responses.

Motivated by this capability and the limitations of conventional feature-based integration, we propose the Beyond-Latent Personalization (BLP) framework, an LLM-enhanced stress classification framework that takes HRV features and LLM-generated domain-specific descriptions as inputs, with each description integrating personal attributes with HRV information from a specific physiological domain. As each HRV domain reflects distinct physiological characteristics, these descriptions provide complementary perspectives on an individual's stress-related physiological responses. To reduce redundancy across their semantic representations, we impose an orthogonality constraint and adaptively aggregate them through learnable source weights, while an intra-subject contrastive objective is employed to enhance stress-discriminative representation learning. This design provides an interpretable alternative to conventional black-box integration: the generated descriptions explicitly reveal how the LLM integrates personal context with physiological information, while the learned source weights quantify the relative contribution of each HRV domain to stress detection.

On TILES-2018, BLP achieves a 29.9\% relative MCC improvement over the strongest baseline, while its performance on WESAD supports its generalizability across datasets. These results demonstrate improved stress classification and transparency into how personal attributes complement physiological information, revealing higher-level relationships less accessible through conventional feature-based integration.

\begin{figure*}[!t]
    \centering
    \includegraphics[width=\textwidth]{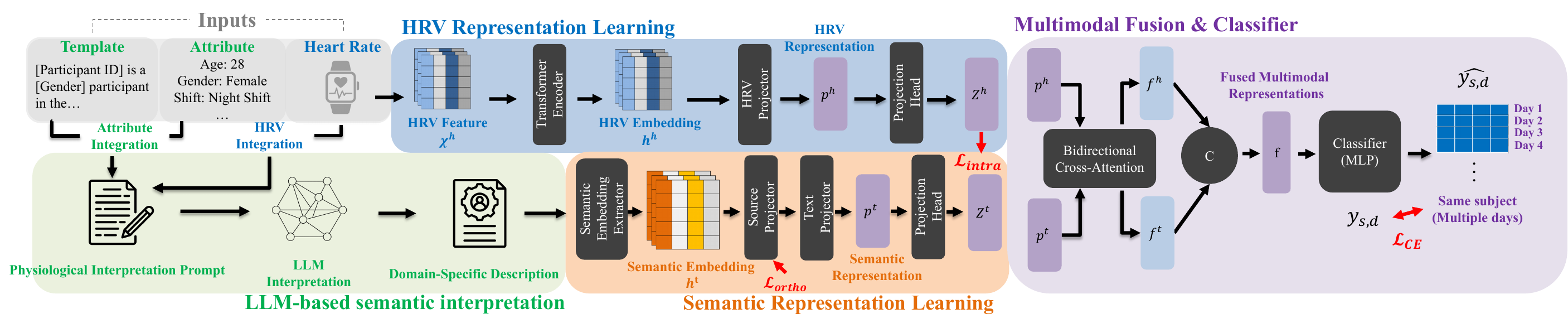}
    \vspace{-7mm}
    \caption{Overview of the proposed BLP framework.}
    \label{fig:framework}
\end{figure*}

\begin{table}[!t]
\centering
\caption{Physiological Interpretation Prompt Template}
\label{tab:prompt}

\scriptsize
\itshape
\renewcommand{\arraystretch}{0.85}
\setlength{\tabcolsep}{2pt}

\begin{tabular}{p{0.96\columnwidth}}
\toprule

\multicolumn{1}{c}{\textbf{\upshape Attribute Integration}}
\\
\midrule

\textbf{\upshape[Neutral Baseline Profile]} \quad
This individual is a \{gender\} participant in the \{age\} age group, and
\{student\_desc\}. Professionally, \{subject\} works as a(n) \{position\},
with \{nurseyears\} of professional experience and \{duration\} of tenure at
the current organization. Working the \{shift\}, \{subject\} averages
\{hours\} per week with approximately \{overtime\} of monthly overtime, and
supervises \{quantsup\}. Regarding personal background, \{subject\} identifies
as \{race\} and has \{englyrs\} of English language experience.
\{possessive\} current relationship status is \{relationship\}, with an annual
household income of \{income\}. \{Subject\} has \{children\} child(ren) under
18 in the household and \{housing\_desc\}. \{possessive\} typical daily
commute is \{commute\_time\}.
\\[1pt]

\midrule
\multicolumn{1}{c}{\textbf{\upshape HRV Integration} (Take time domain as example)}
\\
\midrule

\textbf{\upshape[Data Interpretation Context]} \quad
Note: The following HRV metrics are not instantaneous readings. They are strictly calculated based on rigorous daily aggregation:
Feature Group: 'Time': "Time-domain HRV features (e.g., meanNN, sdNN, RMSSD, pNN50, cvNN)";
Base calculation: Extracted from 5-minute rolling windows of Normal-to-Normal (NN) intervals. Time-domain differences and Frequency domains were normalized;
Suffixes meaning: Metric ending in `\_mean' is the daily average; `\_std' is the daily standard deviation; `\_cv' (Coefficient of Variation) shows daily volatility; `\_skew'/`\_kurt' show distribution shape; `\_1quartile'/`\_3quartile' are the 25th/75th percentiles.
Please interpret these metrics as the overall autonomic stability and allostatic load for the entire day.
\\[1pt]

{\raggedright
\textbf{\upshape[Current Physiological Markers]} \quad
\{feature\_1\}: \{value\_1\} \{feature\_2\}: \{value\_2\}... \{feature\_N\}: \{value\_N\}
\par}
\\[1pt]

\midrule
\multicolumn{1}{c}{\textbf{\upshape General}}
\\
\midrule
\textbf{\upshape[Task]} \quad
Based on the individual's baseline profile and today's HRV metrics, write a daily analytical log assessing their current stress state.
\\[1pt]

\textbf{\upshape[Output Constraints]} \quad
Combine your final analysis into a SINGLE, cohesive paragraph;
Focus strictly on physiological and psychological interpretation;
DO NOT directly state, conclude, or predict whether the individual is ``stressed'' or ``not stressed'';
DO NOT list, quote, or repeat the raw HRV numbers provided above. Use qualitative descriptions (e.g., ``highly volatile,'' ``elevated parasympathetic activity'') instead;
If there's `Unknown' in the description, it means the data is missing and you can ignore it.
\\

\bottomrule
\end{tabular}
\end{table}
\vspace{-4mm}
\section{METHODOLOGY}
\vspace{-2mm}
As illustrated in Fig.~\ref{fig:framework}, BLP consists of four blocks: HRV representation learning, LLM-based semantic interpretation, semantic representation learning, and multimodal fusion. The learned representations from multiple daily samples of the same subject are jointly processed through the bidirectional cross-attention fusion block to capture complementary physiological and semantic information, then concatenated and classified by an MLP for stress detection.
\vspace{-2mm}
\subsection{HRV Representation Learning}
\vspace{-2mm}
For each subject, HRV feature vectors $x^h$ are temporally ordered to form a sequence that captures the evolution of physiological responses over time. Positional encodings are added to $x^h$, which are then processed by a Transformer encoder to model temporal dependencies, yielding encoded HRV embeddings $h^h$. An HRV projector then maps $h^h$ to $p^h$ for bidirectional cross-attention.


\vspace{-3mm}
\subsection{LLM-based Semantic Interpretation}
\vspace{-2mm}
The semantic interpretation module transforms structured physiological and personal information into a textual representation. Following prior studies that contextualize physiological measurements with domain knowledge and user information for LLM-based health analysis \cite{liu2024large, kim2024health}, we convert each sample's HRV features and personal attributes into textual components and incorporate them into predefined \textbf{Physiological Interpretation Prompts}. Specifically, the HRV features are organized into four physiological domains: time, frequency, multi-scale RR, and multi-scale dRR, with a separate prompt constructed for each domain. As shown in Table~\ref{tab:prompt}, the subject's personal attributes are converted into the [Neutral Baseline Profile], which is shared across the four domain-specific prompts, while the HRV features and their interpretation context for each domain are organized into the [Data Interpretation Context] and [Current Physiological Markers]. Together with the predefined [Task] and [Output Constraints], these sections form the complete prompt. During \textbf{LLM Interpretation}, the prompt is processed by an LLM to generate a \textbf{Domain-Specific Description} that describes the physiological state in the context of the individual's baseline characteristics without directly predicting the stress label.
\vspace{-3mm}
\subsection{Semantic Representation Learning}
\vspace{-2.2mm}
The Domain-Specific Descriptions generated by Qwen3.5-9B \cite{qwen3.5} are encoded using Qwen3-Embedding-0.6B \cite{qwen3embedding}, yielding four 1024-dimensional semantic embeddings $h^t$ for each sample. Each embedding is then transformed by a domain-specific Source Projector, producing four learnable representations on which the orthogonal loss is applied. These representations are adaptively aggregated using softmax-normalized learnable weights to capture the relative contribution of each HRV domain. Finally, the aggregated representation is projected into a 100-dimensional shared embedding space, yielding the semantic representation $p^t$ for subsequent cross-modal fusion.
\vspace{-2mm}
\subsection{Training Objective}
\vspace{-2mm}
To capture subject-specific stress patterns while discriminating between stress states within each subject, we employ an intra-subject Multi-Similarity Loss $\mathcal{L}_{intra}$ on the HRV and semantic representations:
\vspace{-2mm}
\begin{equation}
\mathcal{L}_{intra}
=
\mathcal{L}_{MS}
\left(
[\mathbf{Z}^{h};\mathbf{Z}^{t}], \mathbf{c}
\right),
\qquad
c_{s,d}=(s,y_{s,d}),
\vspace{-2mm}
\end{equation}
where $\mathbf{Z}^{h}$ and $\mathbf{Z}^{t}$ denote the HRV and semantic representations obtained through their projection heads, respectively, and $c_{s,d}$ is defined by the subject identity $s$ and stress label $y_{s,d}$. To reduce redundancy among the domain-specific semantic representations, we further introduce an orthogonal loss:
\vspace{-2mm}
\begin{equation}
\mathcal{L}_{ortho}
=
\sum_{1\leq p<q\leq4}
\frac{1}{N^2}
\left\|
\left(\widetilde{\mathbf{E}}^{(p)}\right)^{\top}
\widetilde{\mathbf{E}}^{(q)}
\right\|_{F}^{2},
\vspace{-2mm}
\end{equation}

where $\widetilde{\mathbf{E}}^{(p)}$ and $\widetilde{\mathbf{E}}^{(q)}$ denote the representations from two HRV domains, and $N$ denotes the number of valid samples. This constraint encourages different domain-specific representations to preserve complementary semantic information. Finally, the complete framework is optimized using
\vspace{-2mm}
\begin{equation}
\mathcal{L}_{total}
=
\mathcal{L}_{CE}
+
\lambda_{intra}\mathcal{L}_{intra}
+
\lambda_{ortho}\mathcal{L}_{ortho},
\vspace{-2mm}
\end{equation}
where $\mathcal{L}_{CE}$ denotes the weighted cross-entropy loss for stress classification, while $\lambda_{intra}$ and $\lambda_{ortho}$ control the contributions of the two auxiliary objectives. We provide our source code in a GitHub repository\footnote{\url{https://github.com/Justinma0201/BLP}}.
\vspace{-4mm}
\section{Experiment Setup}
\vspace{-2mm}
\subsection{Dataset and Feature Extraction}
\vspace{-2mm}
TILES-2018 \cite{mundnich2020tiles} is a relatively large-scale longitudinal dataset collected from 212 hospital workers over 10 weeks under naturalistic conditions, whereas WESAD \cite{schmidt2018introducing} is a smaller controlled dataset of 15 participants under Baseline, Stress, and Amusement conditions. Together, they provide complementary settings for evaluating personalized stress detection and framework generalizability. Both datasets use the same 61 HRV features across four domains (Table~\ref{tab:hrv_features}). For TILES-2018, following the data quality screening and label construction in \cite{wang2024wild}, 185 participants are retained. HRV features are extracted from 5-min RR segments derived from heart rate measurements, with a 4-min overlap, and aggregated daily using 13 statistical functionals: mean, standard deviation (SD), coefficient of variation (CV), median, minimum, maximum, 1st and 3rd quartiles, skewness, kurtosis, weighted mean, weighted SD, and weighted CV, resulting in 793 features that are reduced to 100 by Recursive Feature Elimination (RFE). WESAD uses 700-Hz ECG signals to extract HRV features from RR intervals using the recommended 60-s windows with a 0.25-s shift; windows with less than 80\% RR coverage are excluded, and each valid window retains its 61-dimensional HRV vector without daily aggregation or RFE. Following \cite{schmidt2018introducing}, Baseline and Amusement are labeled non-stress and Stress as stress. 

The 17 TILES-2018 attributes from self-reported surveys are gender, age group, student status, current position, professional experience, organizational tenure, work shift, weekly working hours, monthly overtime, number of supervisees, race, English-language experience, relationship status, annual household income, number of children under 18, housing status, and daily commute time. The 11 WESAD attributes include age, height, weight, gender, and dominant hand; coffee consumption, sports participation, smoking status, and illness status on the day of data collection; and coffee consumption and smoking within the last hour.
\begin{table}[!t]
    \centering
    \caption{Overview of the 61 HRV Features}
    \label{tab:hrv_features}
    \includegraphics[width=\columnwidth]{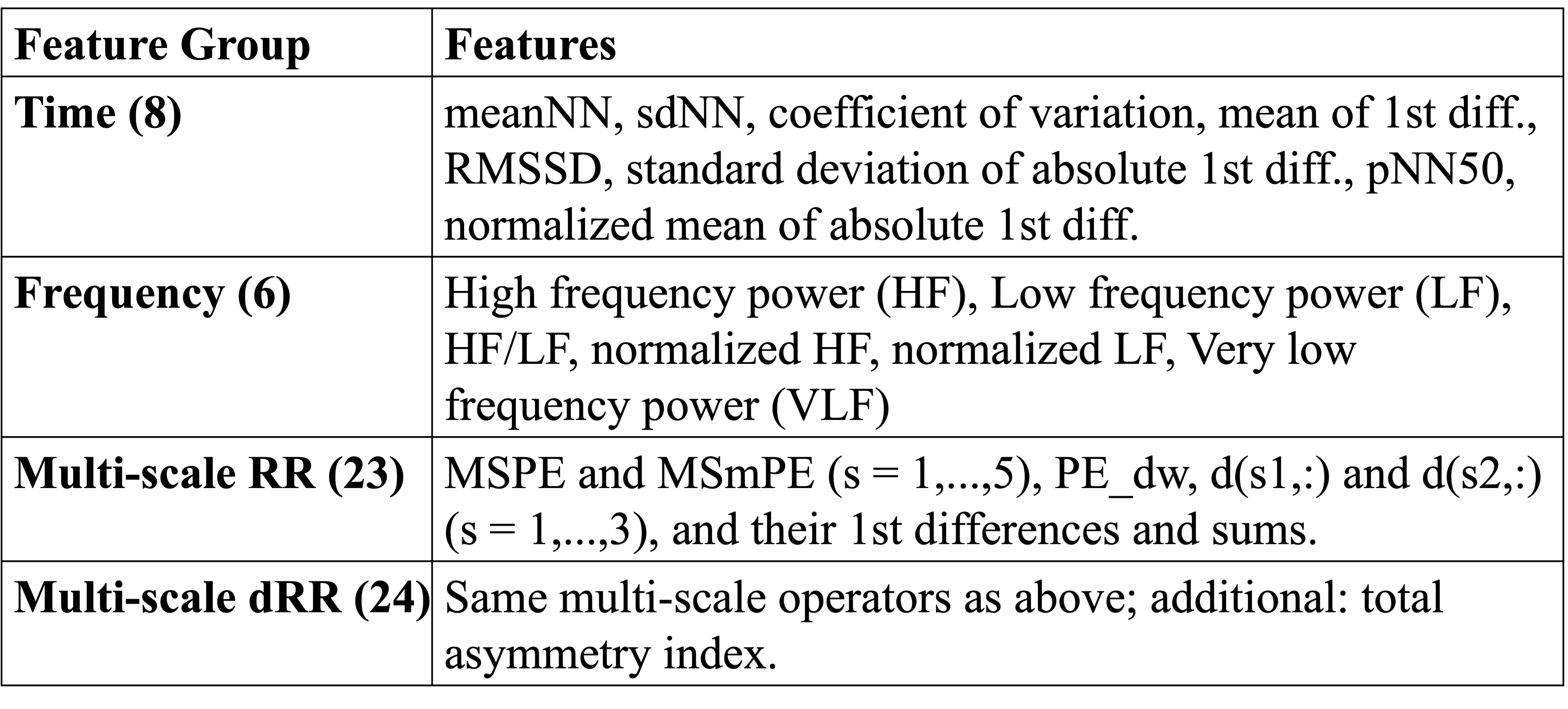}
    \vspace{-5mm}
\end{table}

\begin{table*}[!t]
\centering
\caption{Ablation Results on the TILES-2018 and WESAD Datasets. Interp. = Interpretation; Integ. = Integration; Attr. = Attribute; Extractor = Semantic Embedding Extractor.}
\label{tab:ablation_results}
\resizebox{\textwidth}{!}{
\begin{tabular}{lccccccccccc}
\toprule

\multicolumn{6}{c}{\textbf{Methods}}
& \multicolumn{3}{c}{\textbf{TILES-2018}}
& \multicolumn{3}{c}{\textbf{WESAD}} \\

\cmidrule(lr){1-6}
\cmidrule(lr){7-9}
\cmidrule(lr){10-12}

&
\multirow{2}{*}{\textbf{HRV}}
& \multicolumn{4}{c}{\textbf{Semantic Representation}}
& \multirow{2}{*}{\textbf{UAR}}
& \multirow{2}{*}{\textbf{Acc}}
& \multirow{2}{*}{\textbf{MCC}}
& \multirow{2}{*}{\textbf{UAR}}
& \multirow{2}{*}{\textbf{Acc}}
& \multirow{2}{*}{\textbf{MCC}} \\

\cmidrule(lr){3-6}

&
&
\textbf{LLM Interp.}
& \textbf{Extractor}
& \textbf{HRV Integ.}
& \textbf{Attr. Integ.}
& & & & & & \\

\midrule

LLM-Only
& $\times$ & $\surd$ & $\surd$ & $\surd$ & $\surd$
& 0.516 & 0.504 & 0.040
& 0.616 & 0.683 & 0.260 \\

HRV-Only
& $\surd$ & $\times$ & $\times$ & $\times$ & $\times$
& 0.600 & 0.588 & 0.201
& 0.799 & 0.821 & 0.599 \\

w/o Attribute Integration
& $\surd$ & $\surd$ & $\surd$ & $\surd$ & $\times$
& 0.594 & 0.600 & 0.189
& 0.798 & 0.820 & 0.595 \\

w/o HRV Integration
& $\surd$ & $\surd$ & $\surd$ & $\times$ & $\surd$
& 0.560 & 0.557 & 0.119
& 0.787 & 0.809 & 0.576 \\

w/o Embedding Extractor
& $\surd$ & $\surd$ & $\times$ & $\surd$ & $\surd$
& 0.607 & 0.598 & 0.215
& 0.801 & 0.818 & 0.600 \\

w/o LLM Interpretation
& $\surd$ & $\times$ & $\surd$ & $\surd$ & $\surd$
& 0.619 & 0.610 & 0.240
& 0.799 & 0.815 & 0.595 \\

\midrule

\textbf{Proposed}
& $\surd$ & $\surd$ & $\surd$ & $\surd$ & $\surd$
& \textbf{0.630} & \textbf{0.631} & \textbf{0.261}
& \textbf{0.808} & \textbf{0.823} & \textbf{0.614} \\

\bottomrule
\end{tabular}
}
\end{table*}

\begin{table}[!t]
\vspace{-2mm}
\centering
\caption{Baseline Comparison on TILES-2018 ($^*p<0.05$).}
\label{tab:tiles_comparison}

\footnotesize
\resizebox{\columnwidth}{!}{%
\begin{tabular}{lcccc}
\toprule
\textbf{Method}
& \textbf{UAR}
& \textbf{Acc}
& \textbf{MCC}
& \textbf{MCC $p$-value} \\
\midrule
Trans.+MTL+GRL~\cite{sheikh2022robust}
& 0.584 & 0.603 & 0.172 & 0.0019* \\
\midrule
Trans.+MGCN+MS~\cite{LiCWYX24}
& 0.597 & 0.613 & 0.196 & 0.0085* \\
\midrule
DESHM~\cite{wang2025dynamic}
& 0.598 & 0.619 & 0.201 & 0.0496* \\
\midrule
\textbf{Proposed}
& \textbf{0.630}
& \textbf{0.631}
& \textbf{0.261}
& -- \\
\bottomrule
\end{tabular}%
}
\vspace{1mm}
\raggedright
\end{table}

The HRV encoder uses one Transformer layer with two attention heads and a hidden dimension of 100, while bidirectional cross-attention uses four heads. Training uses a batch size of 16 subjects for up to 400 epochs, with $\lambda_{\mathrm{intra}}$ and $\lambda_{\mathrm{ortho}}$ selected by grid search over $[0.1,1]$. Evaluation uses subject-independent five-fold cross-validation with unweighted average recall (UAR), accuracy (Acc), and MCC, with MCC as the primary metric due to class imbalance. Additional implementation details are provided in the released code.

\vspace{-4mm}
\subsection{Comparison Models}
\vspace{-2mm}
We compare BLP with three baselines and six ablation models. \textbf{Trans.+MTL+GRL}~\cite{sheikh2022robust} incorporates personal attributes through multi-task learning with a gradient reversal layer; \textbf{Trans.+MGCN+MS}~\cite{LiCWYX24} integrates attributes through graph-based modeling with Multi-Similarity loss; and \textbf{DESHM}~\cite{wang2025dynamic} captures higher-order relationships between HRV representations and personal attributes through dynamic hypergraph learning. For the ablations, \textbf{LLM-Only} uses only the semantic representation with a zeroed HRV branch, whereas \textbf{HRV-Only} uses only the HRV representation with a zeroed LLM branch. \textbf{w/o Attribute Integration} and \textbf{w/o HRV Integration} retain both branches but remove personal attributes and HRV information from LLM interpretation, respectively. \textbf{w/o Embedding Extractor} retains the LLM interpretation but removes the subsequent embedding extractor, directly using the normalized last-token representation from the LLM as the semantic embedding. \textbf{w/o LLM Interpretation} removes LLM-based semantic interpretation and directly encodes the original prompt with the embedding extractor.
\vspace{-4mm}
\section{Results Analysis and Ablation}
\vspace{-2mm}
\subsection{Performance Comparison}
\vspace{-2mm}

\textbf{Baselines}: Table~\ref{tab:tiles_comparison} compares the proposed method with three baselines under the same subject-independent setting. The baseline results are reproduced using the original implementation provided by the authors of \cite{wang2025dynamic}. Statistical significance is evaluated using within-fold stratified BCa bootstrap with subject-level paired resampling, with two-sided $p$-values derived from the 95\% CIs of pooled MCC differences and corrected using the Holm--Bonferroni method. The proposed model achieves the highest UAR/Acc/MCC of 0.630/0.631/0.261, with MCC improvements of 51.7\%, 33.2\%, and 29.9\% over Trans.+MTL+GRL, Trans.+MGCN+\allowbreak MS, and DESHM, respectively, all statistically significant ($p<0.05$). These results demonstrate the effectiveness of BLP for personalized stress detection compared with conventional attribute-integration approaches.

\textbf{Ablation Study}: Table~\ref{tab:ablation_results} shows that, on TILES-2018, removing HRV or personal attributes from the LLM prompt reduces MCC by 54.4\% and 27.6\%, respectively. The former shows that HRV provides day-to-day physiological variations and distinct domain information, whose complementarity is preserved by the orthogonal constraint; the latter shows that personal attributes provide subject-specific baselines to contextualize such dynamic information. Together, they enable more complete personalized interpretations. LLM-Only substantially underperforms HRV-Only, indicating that semantics complement rather than replace physiological representations. Removing LLM interpretation reduces MCC by 8.0\%/3.1\% (TILES-2018/WESAD), showing that, although the structured prompt preserves HRV and attribute information, LLM interpretation further establishes semantic relationships between them. Removing the semantic embedding extractor causes larger reductions of 17.6\%/2.3\%, demonstrating its role in extracting task-relevant representations from rich semantics. Its greater impact on TILES-2018 may reflect the more complex daily physiological and personal information in its longitudinal naturalistic setting.
\vspace{-4mm}
\subsection{Case Study}
\vspace{-2mm}

\begin{table}[!t]
\vspace{-2mm}
\centering
\caption{Case Study of HRV-Conditioned LLM Interpretation}
\label{tab:case_study}
\small
\setlength{\tabcolsep}{4pt}

\begin{tabular}{>{\raggedright\arraybackslash}p{\dimexpr\columnwidth-2\tabcolsep\relax}}
\toprule

\textbf{w/o HRV Integration} \textit{(Pred: 0, GT: 1)}:
{\footnotesize
...the cumulative effect of \textcolor{red}{prolonged overtime,
leadership duties, and a lengthy commute} creates a state of
sustained activation where the body and mind are likely operating
under continuous demand, potentially leading to physiological
weariness and psychological strain...
}

\textbf{Proposed} \textit{(Pred: 1, GT: 1)}:
{\footnotesize
...This pattern is further underscored by the
\textcolor{red}{significant variability in very-low-frequency power},
which reflects a disrupted circadian rhythm or allostatic load,
likely exacerbated by the cumulative demands of his
\textcolor{red}{supervisory role, extended overtime hours, and long commute}...
} \\

\midrule

\textbf{w/o HRV Integration} \textit{(Pred: 1, GT: 0)}:
{\footnotesize
...With \textcolor{red}{over 15 years of experience in clinical support
and a tenure of 6 to 9 years at the current organization}, she possesses
significant institutional knowledge and likely established coping mechanisms
... Furthermore, her \textcolor{red}{stable marital status and the presence
of a child under 18} indicate strong social anchoring...
} \\[2pt]

\textbf{Proposed} \textit{(Pred: 0, GT: 0)}:
{\footnotesize
...\textcolor{red}{significantly elevated parasympathetic activity},
as evidenced by the \textcolor{red}{exceptionally high pNN50 and RMSSD values},
which suggest a \textcolor{red}{strong capacity for recovery and stress buffering}... Given her \textcolor{red}{extensive professional tenure and established home environment}, this pattern aligns with a resilient allostatic load profile where the body effectively manages daily demands without exhibiting signs of chronic depletion or dysregulation...
} \\

\bottomrule
\end{tabular}
\end{table}
Table~\ref{tab:case_study} presents cases where BLP corrects prediction errors from the \textit{w/o HRV Integration} setting. Incorporating sample-specific HRV information allows physiological patterns to be interpreted alongside personal attributes. In the stress case, VLF variability is related to occupational demands, whereas in the non-stress case, elevated pNN50 and RMSSD are associated with parasympathetic activity and recovery capacity, with key differences highlighted in red. These cases illustrate how BLP explicitly relates physiological patterns to personal attributes to form individualized, physiologically meaningful representations. This explicit relationship further supports the interpretability of personalized stress detection.

\vspace{-3.75mm}
\section{CONCLUSION and Future Work}
\label{sec:majhead}
\vspace{-2.5mm}
In this work, we proposed BLP, which integrates HRV and personal attributes through LLM-based semantic interpretation to generate personalized, human-readable descriptions fused with temporal HRV representations for stress classification. BLP outperforms all compared baselines on TILES-2018, while its performance on WESAD supports its generalizability across datasets. Ablation results verify the complementary contributions of HRV features, personal attributes, semantic interpretation, and dedicated embedding extraction. However, the framework remains reliant on HRV representations, as LLM-generated semantic representations alone are insufficiently discriminative. Future work will enhance the task-relevant information in LLM-generated descriptions while preserving semantic interpretability, aiming to better balance interpretability and predictive performance.


\newpage
\section{Acknowledgment}
We use OpenAI ChatGPT only to polish manuscript grammar and refine the final delivered code. All experimental results and analyses are conducted without generative AI.

\begingroup

\apptocmd{\thebibliography}{%
  \setlength{\itemsep}{3pt}%
  \setlength{\parsep}{0pt}%
  \setlength{\parskip}{0pt}%
}{}{%
  \PackageWarning{bibspacing}{Failed to patch bibliography}%
}

\bibliographystyle{IEEEbib}
\bibliography{strings,refs}

\endgroup
\end{document}